# A Fuzzy-set-based Joint Distribution Adaptation Method for Regression and its Application to Online Damage Quantification for Structural Digital Twin

Xuan Zhou[a,b], Claudio Sbarufatti[a,1], Marco Giglio[a], Leiting Dong[b,2]

[a] *Department of Mechanical Engineering, Politecnico di Milano, Via La Masa 1, Milano, 20156, Milano, Italy*

[b] *School of Aeronautic Science and Engineering, Beihang University, Xueyuan Road 37, Haidian District, 100191, Beijing, China*

## Abstract

Online damage quantification suffers from insufficient labeled data that weakens its accuracy. In this context, adopting the domain adaptation on historical labeled data from similar structures/damages or simulated digital twin data to assist the current diagnosis task would be beneficial. However, most domain adaptation methods are designed for classification and cannot efficiently address damage quantification, a regression problem with continuous real-valued labels. This study first proposes a novel domain adaptation method, the Online Fuzzy-set-based Joint Distribution Adaptation for Regression, to address this challenge. By converting the continuous real-valued labels to fuzzy class labels via fuzzy sets, the marginal and conditional distribution discrepancy are simultaneously measured to achieve the domain adaptation for the damage quantification task. Thanks to the superiority of the proposed method, a state-of-the-art online damage quantification framework based on domain adaptation is presented. Finally, the framework has been comprehensively demonstrated with a damaged helicopter panel, in which three types of damage domain adaptations (across different damage locations, across different damage types, and from simulation to experiment) are all conducted, proving the accuracy of damage quantification can be significantly improved in a realistic environment. It is expected that the proposed approach to be applied to the fleet-level digital twin considering the individual differences.

[1] Corresponding author. *E-mail address*: claudio.sbarufatti@polimi.it

[2] Corresponding author. *E-mail address*: ltdong@buaa.edu.cn

# 1. Introduction

Ensuring the structural integrity of operating aeronautical and mechanical systems is a critical research topic [1–3]. The digital twin (DT) philosophy, which aims to create a digital representation of the corresponding physical product during the life cycle to support decision-making for operation and sustainment [4], is a potential solution and has drawn much attention in recent years [5–7]. Online damage quantification of structural components through sensor data, as a critical task of structural health monitoring (SHM)[8], is an essential procedure that could be adopted to track and update the state of the digital twin[9].

In a fleet, the damage evolution experienced by each component varies due to differences in manufacture and usage [6]. It is desired to construct an individual damage quantification model for each component. However, acquiring enough labeled data in an engineering environment to train the model is difficult and expensive during the online stage [10], a concern that affects all supervised learning tasks. Since numerous models and data have been accumulated in past applications via experiments or simulations, it might be feasible to reinforce the current damage quantification task with additional labeled data from similar structures/damages or simulations. However, because of the problem-dependent nature of the data and models, adapting them to new structures or damage is challenging.

Domain adaptation methods could be utilized to address the concerns mentioned above. The domain adaptation assumes that labeled data available in a source domain could help improve the task (classification or regression) in an unlabeled (or partially-labeled) target domain by mapping the two domains onto a common latent space on which the data distributions coincide [11]. The more information is shared between two domains, the easier it is to transfer knowledge between them; otherwise, it is more complicated, and even a "negative transfer" may occur [12]. A number of methods [13–16] have been developed for domain adaptation, of which the Transfer Component Analysis (TCA) [13] and Joint Distribution Adaptation (JDA) [14] are two representative methods. TCA evaluates the marginal distribution difference in two domains, while JDA considers the marginal and conditional distribution discrepancies simultaneously and outperforms TCA in most circumstances.

However, current applications in the DT and SHM fields are mainly focused on damage detection, which is a classification problem. Few concerns are paid to the damage quantification as a regression problem. Regarding damage detection, Chakraborty et al. [17] applied the translated inductive transfer learning method to the structural damage classification of an aluminum lug joint. Buisman et al. [18] employed the Feature Alignment Neural Network to transfer SHM data with different boundary conditions in composite structures such that the delamination could be detected accurately with various boundary conditions. Zhang et al. [19] integrated the adaptation-regularization-based transfer learning (ARTL) into the Bayesian model updating (BMU) to bridge the gap between biased numerical models and real structures for damage identification. Omori Yano et al. [20] use the TCA to improve the accuracy of the damage detection. Worden and his co-workers [21–23] published a series of papers on the application of domain adaptation in structural health monitoring, especially for vibration-based damage detection, named Population-Based Structural Health Monitoring (PBSHM). Some deep-learning-based domain adaptation methods [24,25] are also adopted for the damage detection.

Domain adaptations are also used for the unsupervised damage detection [26,27]. However, for the damage quantification, only a few works are reported. In [20] the damage quantification is also invested by combining the TCA and Gaussian process regression (GPR), but the conditional distribution discrepancy is not considered. some methods [28,29] are vision-based similar to image recognition in computer science, which is not the topic in this study. Additionally, limited number of labeled instances is still a key issue for deep learning applications in SHM [30], so exploring damage quantification based on traditional regression methods is still of great importance.

Thus, novel domain adaptation methods for regression are desired for the online damage quantification problem. However, two gaps that limit the application of domain adaptation in damage quantification should be addressed. First, in the damage quantification problem, the response (features) of damage sizes (labels) in different locations or structures has significant variations, indicating different conditional distributions. Therefore, traditional domain adaptation methods such as TCA, which can be used for regression if only considering marginal distribution adaptation, might only work well for adaptation with very similar damages, which is confirmed in Section V. Second, because the label is continuous real-value instead of discrete one-hot, methods that also considering conditional distribution discrepancy, such as JDA, cannot be directly applied due to the difficulty of calculating conditional distribution. To tackle this issue, Wu et al. [31] proposed the online weighted adaptation regularization for regression and applied it to the driver drowsiness estimation from Electroencephalography (EEG) signals, in which fuzzy sets (FSs) are used to transform the continuous label into several fuzzy classes. Because the adaptation matrix and weight matrix of the linear regression are the same, it cannot be applied to nonlinear problems. Liu et al. [32] proposed the conditional distribution deep adaptation regression (CDAR) to tackle the conditional distribution discrepancy in machine health monitoring systems, which only consider the conditional distribution discrepancy. Additionally, this method is deep learning-based and might be challenging to tackle with insufficient samples.

In this study, the Online Fuzzy-set-based Joint Distribution Adaptation for Regression (OFJDAR) is proposed, and an online damage quantification framework with domain adaptation is established. A typical damaged helicopter panel is used to illustrate the method and framework. Results show that the proposed methods could significantly improve prediction accuracy compared with a list of baseline methods. The main contributions of this paper are summarized as follows.

1. In OFJDAR, the regression problem is converted to a classification problem using fuzzy sets inspired by [31] so that the marginal and conditional distributions can be adapted simultaneously, which can achieve better performance compared to the currently adopted TCA method in the SHM field. Also, the processes of adaptation and regression are separated, making it flexible in selecting the proper regressor compared to [31], especially for nonlinear problems.

2. Based on the OFJDAR, an online damage quantification framework for the structural component digital twin with domain adaptation is proposed. It can be easily implemented to improve the accuracy of damage quantification.

3. Three types of domain adaptation in the damage quantification, including different locations of damage, different types of damage, and simulated vs. true damage, are carried out, and the results are compared with several baselines and discussed in detail.

The remainder of this paper is arranged as follows: Section 2 describes the proposed domain adaptation method in detail. Section 3 illustrates the online damage quantification framework with domain adaptation. In Section 4, simulated and experimental datasets from a helicopter panel with two types of damage are used to demonstrate the proposed framework, and the results are compared and discussed in Section 5. In Section 6, this study is completed with some concluding remarks.

## 2. Online Fuzzy-set-based Joint Distribution Adaptation for Regression

This section presents the proposed Online Fuzzy-set-based Joint Distribution Adaptation for Regression (OFJDAR) in detail, which is designed for damage quantification in this study but might be applicable to other similar problems as well.

### 2.1. Problem Definition

We begin with the definitions of terminologies in the domain adaptation. For clarity, the frequently used notations are summarized in Table 1.

In the context of transfer learning, a domain $\mathcal{D}$ involves a d-dimensional feature space $\mathcal{X}$ and a marginal probability distribution $P(\mathbf{x})$, i.e., $\mathcal{D} = \{\mathcal{X}, P(\mathbf{x})\}$, where $\mathbf{x} = \{\mathbf{x}_i\}_{i=1}^{N} \in \mathcal{X}$ is a finite sample from $\mathcal{X}$. Given domain $\mathcal{D}$, a task $\mathcal{T}$ involves a C-cardinality or real-valued label space $\mathcal{Y}$ and a label predictive function $f(\mathbf{x})$, a classifier or regressor, i.e., $\mathcal{T} = \{\mathcal{Y}, f(\mathbf{x})\}$, where $\mathbf{y} \in \mathcal{Y}$, and $f(\mathbf{x}) = Q(\mathbf{y}|\mathbf{x})$ can be explained as the conditional probability distribution.

Table 1 Notations and descriptions used in this paper.

| Notation | Description |
| --- | --- |
| $\mathbf{A}$ | adaptation matrix |
| $f$ | adapted regressor |
| $\mathbf{H}$ | center matrix |
| $\lambda$ | regularization parameter |
| $m, C$ | number of shared features/classes |
| $k$ | number of subspace bases |
| $\mathcal{D}^s, \mathcal{D}^t$ | source/target domain |
| $\mathbf{K}$ | kernel matrix |
| $\mathbf{M}_{(c)}$ | MMD matrices, $c \in 0, \ldots, C$ |
| $\widetilde{\mathbf{M}}_{(c)}$ | Fuzzy edition of MMD matrices, $c \in 0, \ldots, C$ |
| $n_s, n_t$ | number of source/target samples |
| $P, Q$ | marginal/conditional probability distribution |
| $\mu$ | membership degree matrix |
| $\mathbf{X}$ | Input matrix |
| $\mathbf{x}^s$ | labeled features in the source domain |
| $\mathbf{x}^{tl}$ | labeled samples in the target domain |
| $\mathbf{x}^{tu}$ | unlabeled samples in the target domain |
| $\mathbf{y}^s$ | labels of samples in the target domain |
| $\mathbf{x}^{tl}$ | labels of samples in the target domain |
| $\hat{\mathbf{y}}^{tu}$ | predicted labels of unlabeled samples in the target domain |

Given a source domain $\mathcal{D}^s$ with corresponding task $\mathcal{T}^s$, and a target domain $\mathcal{D}^t$ with corresponding task $\mathcal{T}^t$ which is unlabeled or partial-labeled. Domain adaptation is defined as the process of improving the label predictive function $f^t(\mathbf{x})$ by using the related information from $\mathcal{D}^s$ and $\mathcal{T}^s$, where $\mathcal{D}^s \neq \mathcal{D}^t$ ( $\mathcal{X}^s \neq \mathcal{X}^t$ or $P^s(\mathbf{x}^s) \neq P^t(\mathbf{x}^s)$ ) or $\mathcal{T}^s \neq \mathcal{T}^t$ ( $\mathcal{Y}^s \neq \mathcal{Y}^t$ or $Q^s(\mathbf{y}^s|\mathbf{x}^s) \neq Q^t(\mathbf{y}^t|\mathbf{x}^t)$))

Typical domain adaptation methods minimize the distribution difference between the source and target domain by migrating features to each other using feature transformation or transforming the features of the source and target domain into a common feature space [11]. Then, traditional machine learning methods could be used for the classification or regression problem. The core of Transfer Component Analysis (TCA) [13] is to use Maximum Mean Discrepancy (MMD) [33] as a metric to minimize the discrepancy of marginal distributions $P^s(\mathbf{x}^s) \neq P^t(\mathbf{x}^s)$. Furthermore, if the conditional distribution of the two domains is also different, i.e., $Q^s(\mathbf{y}^s|\mathbf{x}^s) \neq Q^t(\mathbf{y}^t|\mathbf{x}^t)$, simultaneously reducing the marginal and conditional distribution differences is required, which could be implemented by the Joint Distribution Adaptation [14], which aims to learn a feature representation in which the distribution differences between 1) $P^s(\mathbf{x}^s)$ and $P^t(\mathbf{x}^s)$, 2) $Q^s(\mathbf{y}^s|\mathbf{x}^s)$ and $Q^t(\mathbf{y}^t|\mathbf{x}^t)$ are explicitly reduced.

In this study, due to the damage growing progressively in the online stage and the monitoring data being acquired sequentially, the online damage quantification problem could be considered an Online Joint Distribution Adaptation for Regression problem, as defined in Problem 1 and shown in Figure 1. However, there is a challenge to solve this problem by JDA. The primary JDA is designed to process classification problems, as seen from its definition. In its implementation, the conditional distribution $Q(\mathbf{y}|\mathbf{x})$ is approximated by the class conditional probability distribution $Q(\mathbf{x}|\mathbf{y})$, where the detail is introduced in Section 2.3. Nevertheless, it would be difficult for regression problems to define the $Q(\mathbf{x}|\mathbf{y})$ since these are no class but continuous real-valued labels. Therefore, some modifications of the primary JDA method are required to tackle this challenge, which is the motivation of the proposed method.

**Problem 1.** *(Online Joint Distribution Adaptation for Regression) Given the labeled source domain with real-value label $\mathcal{D}^s = \{(\mathbf{x}_1^s, y_1^s), \ldots, (\mathbf{x}_{n_s}^s, y_{n_s}^s)\}$, and the target domain with $n_{tl}$ calibrated (labeled) samples $\{\mathbf{x}_1^t, \ldots, \mathbf{x}_{n_{tl}}^t\}$ and $n_{tu}$ uncalibrated (unlabeled) samples $\{\mathbf{x}_{n_{tl}+1}^t, \ldots, \mathbf{x}_{n_{tl}+tu}^t\}$ acquired sequentially, under the assumptions that $\mathcal{X}^s = \mathcal{X}^t$, $\mathcal{Y}^s = \mathcal{Y}^t$, $P^s(\mathbf{x}^s) \neq P^t(\mathbf{x}^t)$, $Q^s(\mathbf{y}^s|\mathbf{x}^s) \neq Q^t(y^t|\mathbf{x}^t)$. Online Joint Distribution Adaptation for Regression aims to incrementally learn a feature representation in which the distribution differences between 1) $P^s(\mathbf{x}^s)$ and $P^t(\mathbf{x}^t)$, 2) $Q^s(\mathbf{y}^s|\mathbf{x}^s)$ and $Q^t(\mathbf{y}^t|\mathbf{x}^t)$ are explicitly reduced.*

The core of the Online Joint Distribution Adaptation for Regression is to find an orthogonal transformation matrix **A** incrementally which could project the data to a $k$-dimensional space, in which the distance between the transformed $P(\mathbf{A}^T\mathbf{x}^s)$ and $P(\mathbf{A}^T\mathbf{x}^t)$ and the distance between $Q(\mathbf{y}^s|\mathbf{A}^T\mathbf{x}^s)$ and $Q(\mathbf{y}^t|\mathbf{A}^T\mathbf{x}^t)$ are minimized simultaneously. Naturally, two types of adaptations are required: the marginal distribution adaptation and conditional distribution adaptation.

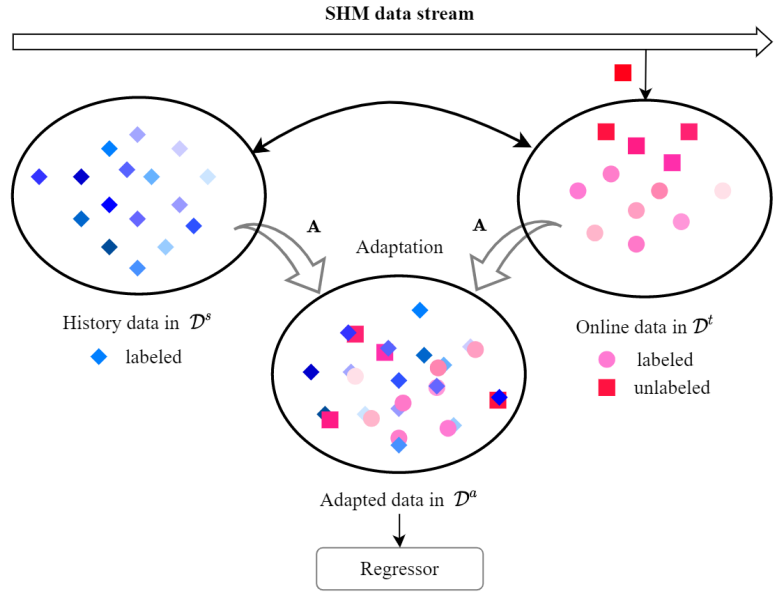

Figure 1 Utilize the historical labeled data in similar damages and available labeled data during online monitoring by domain adaptation methods to improve the prediction of unlabeled sensor data sequentially for the damage quantification (a regression) problem.

### 2.2. Marginal Distribution Adaptation

For different damage types/locations, or between the simulated and realistic data, the marginal distributions of the sensor data would be different. The marginal distribution adaptation adopted in this study is the same as that in TCA and JDA. Thus only a brief introduction is given here, and the interested reader could refer to [13,14] for more details. To reduce the difference between marginal distributions $P^s(\mathbf{x}^s)$ and $P^t(\mathbf{x}^t)$, the empirical Maximum Mean Discrepancy (MMD) is adopted as the distance measure to compare different distributions. Specifically, MMD computes the distance between the sample means of the source and target data by:

$$\left\| \frac{1}{n_s} \sum_{\mathbf{x}_i \in \mathcal{D}^s} \mathbf{A}^T \phi(\mathbf{x}_i) - \frac{1}{n_t} \sum_{\mathbf{x}_j \in \mathcal{D}^t} \mathbf{A}^T \phi(\mathbf{x}_j) \right\|^2 = \text{tr}(\mathbf{A}^T \mathbf{K} \mathbf{M}_0 \mathbf{K}^T \mathbf{A}) \tag{1}$$

where $\phi$ is the RBF (Radial Basis Function) kernel function, and **K** is the RBF (Radial Basis Function) kernel matrix, which is adapted here to map the original data to a higher dimensional space and is computed by:

$$K(\mathbf{x}, \mathbf{x}') = \exp\left(-\frac{\|\mathbf{x} - \mathbf{x}'\|^2}{2\sigma^2}\right) = \exp(-\gamma \|\mathbf{x} - \mathbf{x}'\|^2) \tag{2}$$

where $\|\mathbf{x} - \mathbf{x}'\|^2$ may be recognized as the squared Euclidean distance between the two feature vectors, and $\gamma = \frac{1}{2\sigma^2}$. It is worth mentioning that $\gamma$ is an important parameter in this algorithm. Intuitively, a small $\gamma$ defines a Gaussian function with large variance, in which case two samples could be considered similar even if they are far from each other.

**M**$_0$ is the MMD matrix and is computed as follows

$$(M_0)_{ij} = \begin{cases} \dfrac{1}{n_s n_s}, & \mathbf{x}_i, \mathbf{x}_j \in \mathcal{D}^s \\ \dfrac{1}{n_t n_t}, & \mathbf{x}_i, \mathbf{x}_j \in \mathcal{D}^t \\ \dfrac{-1}{n_s n_t}, & \text{otherwise} \end{cases} \tag{3}$$

## 2.3. Conditional Distribution Adaptation by converting the regression task to a classification task

For the damage quantification in SHM, the damage location and damage type would affect the response signal of the sensors. The modeling bias would also introduce a different response between the simulated model and the realistic structure. That means, even with the same damage size, for different domains, the sensor response would also be different, which can be considered the conditional distribution discrepancies. In this circumstance, only reducing the difference in marginal distributions of two domains is insufficient. However, two challenges need to be addressed here to consider the difference in conditional distributions.

First, since these are unlabeled data in the target domain, the conditional distribution $Q^t(\mathbf{y}^{tu}|\mathbf{x}^{tu})$ cannot be directly modeled. Following [14], the pseudo labels of the unlabeled target data is computed first with a basic classifier, and the class-conditional distributions $Q^s(\mathbf{x}^s|\mathbf{y}^s))$ and $Q^t(\mathbf{x}^t|\mathbf{y}^t))$ are used instead of the conditional distribution $Q^s(\mathbf{y}^s|\mathbf{x}^s)$ and $Q^t(\mathbf{y}^t|\mathbf{x}^t)$, respectively. Afterward, an analogous way as described in Section 2.2 is used to compute the MMD of the conditional distributions between the source and target domain.

$$\left\| \frac{1}{n_s^{(c)}} \sum_{\mathbf{x}_i \in \mathcal{D}_{(c)}^s} \mathbf{A}^\mathrm{T} \phi(\mathbf{x}_i) - \frac{1}{n_t^{(c)}} \sum_{\mathbf{x}_j \in \mathcal{D}_{(c)}^t} \mathbf{A}^\mathrm{T} \phi(\mathbf{x}_j) \right\|^2 = \mathrm{tr}(\mathbf{A}^\mathrm{T} \mathbf{K} \mathbf{M}_c \mathbf{K}^\mathrm{T} \mathbf{A}) \tag{4}$$

where $\mathcal{D}_{(c)}^s = \{\mathbf{x}_i : \mathbf{x}_i \in \mathcal{D}^s \land y(\mathbf{x}_i) = c\}$ is the set of samples belonging to class $c$ in the source domain, $y(\mathbf{x}_i)$ is the true label of $\mathbf{x}_i$, and $n_{(c)}^s = |\mathcal{D}_{(c)}^s|$. Correspondingly, $\mathcal{D}_{(c)}^t = \{\mathbf{x}_i : \mathbf{x}_i \in \mathcal{D}^t \land y(\mathbf{x}_i) = c\}$ is the set of samples belonging to class $c$ in the target domain, $y(\mathbf{x}_j)$ is the pseudo (predicted) label of $\mathbf{x}_j$, and $n_{(c)}^t = |\mathcal{D}_{(c)}^t|$. Thus, the MMD matrices $\mathbf{M}_c$ involving class labels are computed as follows:

$$(\mathbf{M}_c)_{ij} = \begin{cases} \dfrac{1}{n_s^{(c)} n_s^{(c)}}, & \mathbf{x}_i, \mathbf{x}_j \in \mathcal{D}_{(c)}^s \\ \dfrac{1}{n_t^{(c)} n_t^{(c)}}, & \mathbf{x}_i, \mathbf{x}_j \in \mathcal{D}_{(c)}^t \\ \dfrac{-1}{n_s^{(c)} n_t^{(c)}}, & \begin{cases} \mathbf{x}_i \in \mathcal{D}_{(c)}^s, \mathbf{x}_j \in \mathcal{D}_{(c)}^t \\ \mathbf{x}_j \in \mathcal{D}_{(c)}^s, \mathbf{x}_i \in \mathcal{D}_{(c)}^t \end{cases} \\ 0, & \text{otherwise} \end{cases} \tag{5}$$

Second, in the regression task, these are continuous real-valued labels instead of discrete one-hot ones, implying the MMD matrices $\mathbf{M}_c$ cannot be directly computed since these are no class labels. In this study, a fuzzy-set-based method [31] is adopted to tackle this challenge. The main idea of this method is to transform the continuous real-value label to the fuzzy membership degree, which would then be adopted to compute the fuzzy edition of the MMD matrix $\widetilde{\mathbf{M}}_{(c)}$, which is similar to Eq. (5).

### 2.3.1. Computing the membership degree of samples with Fuzzy sets

Inspired by [31], a fuzzy-set-based method is chosen to obtain the membership degree of samples, which converts the real-valued damage sizes into fuzzy class labels, i.e., assigning each sample to a certain class according to its affiliation. Then the approximate calculation of class conditional probabilities $Q^s(\mathbf{x}^s|\mathbf{y}^s))$ and $Q^t(\mathbf{x}^t|\mathbf{y}^t))$ is calculated, and the conditional distribution discrepancy can be measured to some degree, which allows for domain adaptation using the philosophy of the JDA. It should be noted that the conversion of the labels is only used to calculate the conditional distribution discrepancy. When training the damage quantification model with the adapted data, the real-valued labels are still used.

The pseudo-code about calculating the membership degree by Fuzzy Set is shown in Algorithm 1.

**Algorithm 1** Calculate the membership degree by Fuzzy Set.

**Input:** labels **y** with real value, number of fuzzy sets $n_{fs}$, percentile values.
**Output:** Membership degree $\mu$
1: Fit the distribution of labels $\hat{f}_h(\mathbf{y})$ by Kernel Density Estimation (KDE);
2: Calculate the percentiles by integral of the distribution $\hat{f}_h(\mathbf{y})$, typically $p_5, p_{50}$ and $p_{95}$;
3: Generate $C$ fuzzy set based on the percentile values.
4: Calculate the membership degree **u**.

Taken three fuzzy sets with triangular membership functions as an example. First, for the $n$ continuous labels $\{y_i\}_{i=1,\dots,n}$, which have been scaled to [0,1]. the 5, 50, and 95 percentiles $p_5, p_{50}, p_{95}$, are computed by Kernel Density Estimation (KDE), respectively, and three triangular FSs, Small, Medium, and Large, are defined based on them, as shown in Figure 2. In this way, the real-valued labels can be "classified" into three fuzzy classes, corresponding to the different classes in a traditional classification problem. It is worth mentioning that, in the case of fuzzy classes, a sample could be assigned to multiple classes simultaneously, at varying degrees, which is different from the traditional classification problem.

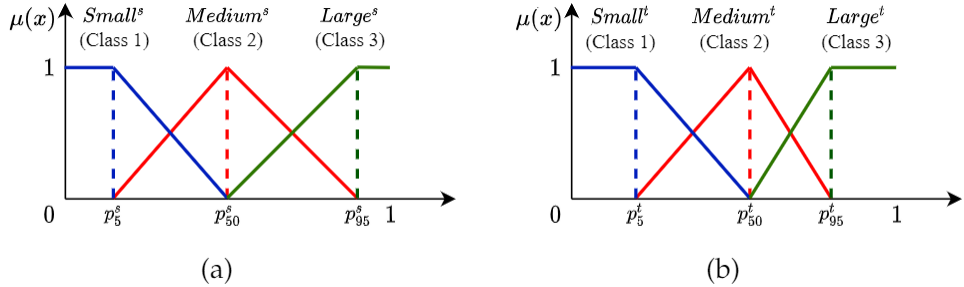

Figure 2 Three FSs with triangular membership function in the (a) source domain, and (b) target domain. Adopted from [31].

### 2.3.2. Constructing the MMD matrices $\widetilde{\mathbf{M}}_{(c)}$ incorporating fuzzy membership degree

With the computed membership degree in Section 2.3, by analogy with Eq. (5), the MMD of the conditional distribution could be computed as follows in this study:

$$\left\| \sum_{\mathbf{x}_i \in \mathcal{D}_s} \bar{\mu}_{ic}^s \mathbf{A}^T \phi(\mathbf{x}_i) - \sum_{\mathbf{x}_j \in \mathcal{D}_t} \bar{\mu}_{ic}^t \mathbf{A}^T \phi(\mathbf{x}_j) \right\|^2 = \text{tr}(\mathbf{A}^T \mathbf{K} \widetilde{\mathbf{M}}_{(c)} \mathbf{K}^T \mathbf{A}) \tag{6}$$

where $\widetilde{\mathbf{M}}_{(c)}$ is the fuzzy edition of $\mathbf{M}_{(c)}$, which is computed as follows:

$$(\widetilde{M}_{(c)})_{ij} = \begin{cases} \bar{\mu}_{ic}^s \bar{\mu}_{jc}^s, & x_i, x_j \in \mathcal{D}^s \\ \bar{\mu}_{ic}^t \bar{\mu}_{jc}^t, & x_i, x_j \in \mathcal{D}^t \\ -\bar{\mu}_{ic}^s \bar{\mu}_{jc}^t, & \begin{cases} x_i \in \mathcal{D}^s, x_j \in \mathcal{D}^t \\ x_j \in \mathcal{D}^s, x_i \in \mathcal{D}^t \end{cases} \\ 0, & \text{otherwise} \end{cases} \quad (7)$$

where $\bar{\mu}_{ic}$ is normalized according to its class as follow:

$$\bar{\mu}_{ic} = \frac{\mu_{ic}}{\sum_{i=1}^{n} \mu_{ic}}, \quad i = 1, \dots, n; \ c = 1, \dots, C. \quad (8)$$

It is worth noting that in Eq. (7) all samples participate in the computation of $\widetilde{M}_{(c)}$, which is different from Eq. 4 in primal JDA.

## 2.4. Finding the adaptation matrix by solving a constrained optimization problem

After computing $\widetilde{M}_{(c)}$ as in Section 2.3, the following constrained optimization is adopted to find the optimal adaptation matrix $A$. The procedure is the same as that in the classical JDA method and the interested reader can refer to [14] for more detail.

The constrained optimization problem is expressed as follows:

$$\min_{A^T K H K^T A = I} \sum_{c=0}^{C} \text{tr}\left(A^T K \widetilde{M}_{(c)} K^T A\right) + \lambda \parallel A \parallel_F^2 \quad (9)$$

where $A^T K H K^T A = I$ is a constraint that the variance of the data before and after the transformation has to remain the same, $H$ is the central matrix, and $I$ is the unit matrix. $\widetilde{M}_{(0)}$ is the MMD matrix of the marginal distribution, which is the same as that in primal JDA, thus $\widetilde{M}_{(0)} = M_0$, $\lambda$ is the regularization parameter to guarantee the optimization problem to be well-defined.

The constrained optimization in Eq. (9) could be solved by constructing its Lagrange function:

$$\begin{aligned} L &= \text{tr}\left(A^T \left(X \sum_{c=0}^{C} \widetilde{M}_{(c)} X^T + \lambda I\right) A\right) \\ &+ \text{tr}\left((I - A^T X H X^T A)\Phi\right) \end{aligned} \quad (10)$$

where $\Phi = \text{diag}(\phi_1, \dots, \phi_k) \in \mathbb{R}^{k \times k}$ is the Lagrange multiplier.

Setting $\frac{\partial L}{\partial A} = 0$, we obtain the generalized eigen decomposition:

$$\left(X \sum_{c=0}^{C} \widetilde{M}_{(c)} X^T + \lambda I\right) A = X H X^T A \Phi \quad (11)$$

Finally, the optimal adaptation matrix **A** could be computed by solving Eq. (11) for the $k$ smallest eigenvectors.

## 2.5. OFJDAR Algorithm

The pseudo-code of the complete OFJDAR algorithm is described in Algorithm 2, in which a pseudo label refinement scheme is added. It is worth noting that in OFJDAR, differently from [31], the operation of domain adaptation and regressor training are separated, guaranteeing the flexibility in choosing a suitable regression method.

---

**Algorithm 2** Online Fuzzy-set-based Joint Distribution Adaptation for Regression (OFJDAR)

---

**Input:** $\mathbf{x}^s, \mathbf{x}^{tl}, \mathbf{x}^{tu}, \mathbf{y}^s, \mathbf{y}^{tl}$
**Output:** Adaptation matrix **A**, predicted label $\hat{\mathbf{y}}^{tu}$, Adapted regressor $f$

1: Stack the input data together to construct matrix $\mathbf{X} = [\mathbf{x}^s, \mathbf{x}^{tl}, \mathbf{x}^{tu}]$, $\mathbf{X}^l = [\mathbf{x}^s, \mathbf{x}^{tl}]$, $\mathbf{Y}^l = [\mathbf{y}^s, \mathbf{y}^{tl}]$
2: Construct MMD matrix $\mathbf{M}_0$ for marginal distribution adaptation by Eq. (3);
3: Train a regressor $f$ on $(\mathbf{X}^l, \mathbf{Y}^l)$ by the Gaussian process regression, feed the $\mathbf{x}^{tu}$ as the input of $f$ to predict the pseudo label $\hat{\mathbf{y}}^{tu}$, and stack it with the labels $\mathbf{y}^{tl}$ and $\hat{\mathbf{y}}^{tu}$ to construct $\mathbf{Y} = [\mathbf{y}^s, \mathbf{y}^{tl}, \hat{\mathbf{y}}^{tu}]$;
4: **While** not convergence **do**
5:     Computer the membership degrees $\mu$ by Algorithm 1.
6:     Construct MMD matrices $\{\widetilde{\mathbf{M}}_{(c)}\}_{c=1}^C$ of the conditional distribution adaptation by Eq. (7) and Eq. (8).
7:     Solve the generalized eigen decomposition problem in Eq. (11) (eig in Matlab or scipy.linalg.eig in Python) and select the $k$ smallest eigenvectors to construct the adaptation matrix **A**
8:     Train the regressor $f$ again on $(\mathbf{A}^T\mathbf{X}^l, \mathbf{Y}^l)$, and update the pesudo label $\hat{\mathbf{y}}^{tu}$ and **Y**;
9: **End While**
10: **Return** an adapted regressor $f$, predicted label $\hat{\mathbf{y}}^{tu}$.

---

For the OFJDAR algorithm, the input is the features $\mathbf{x}^s$ and labels $\mathbf{y}^s$ in the source domain, available features $\mathbf{x}^{tl}$ and labels $\mathbf{y}^{tl}$ in the target domain, and current unlabeled features $\mathbf{x}^{tu}$.

In step 1, the matrix **X** contains all features, $\mathbf{X}^l$ contains all labeled features, $\mathbf{Y}^l$ contains all true labels are constructed by stacking the inputs. Then in step 2, the MMD matrix $\mathbf{M}_0$ is calculated by Eq. (3) since the marginal distribution adaptation only relies on features which have been provided. In step 3, the pseudo labels $\hat{\mathbf{y}}^{tu}$ are obtained by training a regressor based on available labeled samples. Even though $\hat{\mathbf{y}}^{tu}$ are pseudo labels, it allows us to construct the matrix **Y**, completing the input of Algorithm 1.

During the loop from step 4 to 9, the procedures follow the order in Section 2.3 and 2.4, and output the regressor $f$ and the predicted labels $\hat{\mathbf{y}}^{tu}$ after adaptation, which is desired to be more accurate than that in step 3. With the supposition that the result could be improved by iteration, a pseudo label refinement scheme is added that procedures in the loop are repeated with the newly predicted labels $\hat{\mathbf{y}}^{tu}$. Once the result convergent, either by the maximum iteration number or $\|\hat{\mathbf{y}}^{tu}_{k+1} - \hat{\mathbf{y}}^{tu}_k\| < \epsilon$, the adapted regressor $f$ and predicted label $\hat{\mathbf{y}}^{tu}$ are outputted as the final result.

Additionally, a simple migration of TCA to be applicable for the online damage quantification problem is also presented here as a baseline method, as shown in Algorithm 3. Although in [20] TCA is combined with GPR to conduct the damage quantification, it is not integrated into an online framework.

---

**Algorithm 3** Online Transfer Component Adaptation for Regression (OT-CAR)

**Input:** $\mathbf{x}^s$, $\mathbf{x}^{tl}$, $\mathbf{x}^{tu}$, $\mathbf{y}^s$, $\mathbf{y}^{tl}$
**Output:** Adaptation matrix **A**, predicted label $\hat{\mathbf{y}}^{tu}$, Adapted regressor $f$
1: Construct matrice $\mathbf{X} = [\mathbf{x}^s, \mathbf{x}^{tl}, \mathbf{x}^{tu}]$, $\mathbf{X}^l = [\mathbf{x}^s, \mathbf{x}^{tl}]$, $\mathbf{Y}^l = [\mathbf{y}^s, \mathbf{y}^{tl}]$
2: Construct MMD matrix $\mathbf{M}_0$ by Eq. (2);
3: Solve the generalized eigen decomposition problem in Eq. (11) where $C = 0$ and select the $k$ smallest eigenvectors to construct the adaptation matrix **A**
4: Train a regressor on $(\mathbf{X}^l, \mathbf{Y}^l)$, predict the pesudo label $\hat{\mathbf{y}}^{tu}$, and constrcut $\mathbf{Y} = [\mathbf{y}^s, \mathbf{y}^{tl}, \hat{\mathbf{y}}^{tu}]$;
5: Train a regressor $f$ on $(\mathbf{A}^T \mathbf{X}^l, \mathbf{Y}^l)$, and predict the label $\hat{\mathbf{y}}^{tu}$;
6: **Return** an adapted regressor $f$, predicted label $\hat{\mathbf{y}}^{tu}$.

---

### 2.6. Optimization of hypermeter $\gamma$

Hypermeter $\gamma$ in the RBF kernel function would significantly affect the adaptation quality. From Eq. (2) we can see that the smaller $\gamma$ is, the larger the distance between two samples will be after transformation. In other words, $\gamma$ defines how far the influence of a single example reaches, with low values meaning 'far' and high values meaning 'close'. In the online stage, since the labels of the unlabeled samples are unavailable, we do not know which $\gamma$ would perform best for the domain adaptation. An inappropriate $\gamma$ may lead to poor predictions. Here, a solution is proposed to address this problem.

The unlabeled samples are closer to the newly obtained labeled samples than the others in the target domain, since the samples are sequentially acquired. Therefore, in optimizing $\gamma$, several new labeled samples are considered temporarily unlabeled and merged with unlabeled samples to form an augmented unlabeled dataset $\widetilde{\mathcal{D}}^{tu}$. Then, domain adaptations with different $\gamma$ are conducted to predict the labels of $\widetilde{\mathcal{D}}^{tu}$, and the accuracy on the selected labeled samples could be evaluated. Finally, $\gamma$ that performs best on $\widetilde{\mathcal{D}}^{tu}$ is then applied to the adaptation on the original $\mathcal{D}^{tu}$ assuming that it could maintain its quality.

## 3. Online Damage Quantification with the Proposed Domain Adaptation Method

In this section, we proposed an online damage quantification framework with domain adaptation. In the framework, the proposed domain adaptation methods are integrated to assist the current online damage quantification task by adapting knowledge from similar structures or damages.

### 3.1. The proposed framework

The online damage quantification framework with domain adaptation is shown in Figure. 3.

In the offline stage, a database $\mathbf{D}^h$ containing all available historical data was constructed by collecting damage quantification data obtained from similar structures/damages or simulations. Prior to the online structural damage quantification, the source domain data $\mathbf{D}^s$ are extracted from $\mathbf{D}^h$ to assist current task, and sensors monitoring the damage state were deployed on the target structure.

In the online stage, to begin the process, sequential monitoring data is collected from the deployed sensors, followed by feature extraction, which produces the unlabeled feature $\mathbf{x}^{tu}$. The proposed domain adaptation methods are then utilized to map the features into a common latent space in conjunction with the labeled historical data from the database. A regressor is trained upon the adapted labeled data. Following the construction of the regressor, the damage size $\hat{\mathbf{y}}^{tu}$ could be easily obtained, which would be utilized to support the decision-making in a realistic environment. Once the true damage size $\mathbf{y}^{tu}$ is measured, the prediction accuracy could be evaluated, and the $\mathbf{y}^{tu}$, paired with the $\mathbf{x}^{tu}$, would constitute new labeled samples that could be stored to enrich the database.

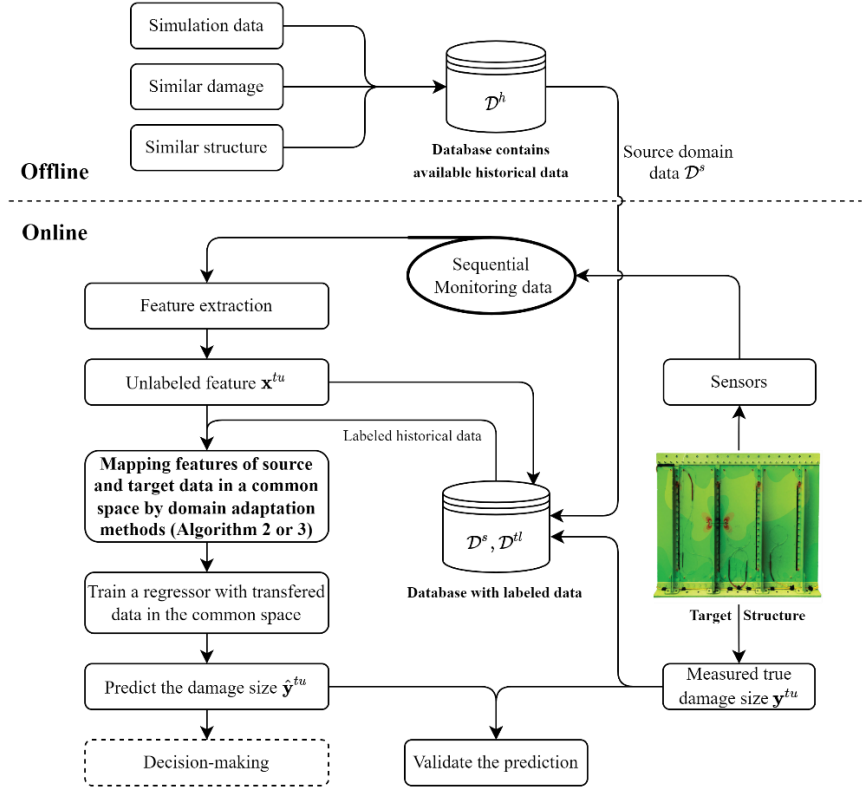

Figure 3 The framework of online damage quantification with domain adaptation.

### 3.2. Extract damage-sensitive features from sensor data

Damage-sensitive features are some quantities extracted from the measured system response data that are used to indicate the presence of damage in a structure [8]. With the arrival sensor data, feature extraction transforms the raw data into some alternative form where the correlation with the damage is more readily observed. The widely used methods include Fourier transform, Wavelet Transform, Empirical Mode Decomposition [34]. Machine learning-based methods, such as Autoregressive [35], Autoencoder [36], also draw much attention.

### 3.3. Mapping features to a new space with proposed domain adaptation methods and predicting the damage size

The labeled data in the database $\mathcal{D}^s$ and $\mathcal{D}^{tl}$, combined with the unlabeled features $\mathbf{x}^{tu}$ extracted from sensors, construct the input of the proposed domain adaptation methods (See Algorithm 2). These data are then mapped to a common latent space, in which an adapted regressor $f$ is trained, and the damage size could be predicted in real-time.

### 3.4. True damage size measurement and storage

Once the true damage size could be measured, by non-destructive inspection (might combine with physics-based model), the true labels $\mathbf{y}^{tu}$, paired with features $\mathbf{x}^{tu}$, will be stored to enrich the database. The accuracy of the prediction could also be evaluated.

## 4. Application to the damage quantification of a helicopter panel

### 4.1. Helicopter panel and its numerical model

In this study, an aluminum alloy stiffened panel [37,38] is considered, as shown in Figure 4, which is representative of the rear fuselage of a medium-heavy weight helicopter. The panel specimen is composed of a skin stiffened with four riveted stringers and some additional reinforcements in the upper and lower edges regions, specifically designed to produce a stress field consistent with that experienced by the real structure by distributing the applied load along the entire specimen width. The skin is 600 mm wide, 500 mm tall, and 0.81 mm thick and is made of Al2024-T6 alloy. The four L-shaped stringers are made of Al7075-T76 alloy, being 435 mm long and 1.2 mm thick and evenly spaced over the specimen skin surface. The panel is equipped with a network of $s = 20$ fiber Bragg grating (FBG) strain gauges located on the stringers as shown in Figure 4(c) and the strain is measured along their length direction.

In this study, two types of damage are considered: one is referred to as "rivet cracking", in which a rivet is removed and then a crack is propagated from an artificial defect located in the panel skin. The other is called "stringer failure", which represents the worst case of damage and corresponds to the maximum crack propagation rate. Stringer failure causes a severe strain field modification due to a sudden change in the load path along the structure. The strain field has been selected as the damage-dependent variable upon which to generate the SHM dataset.

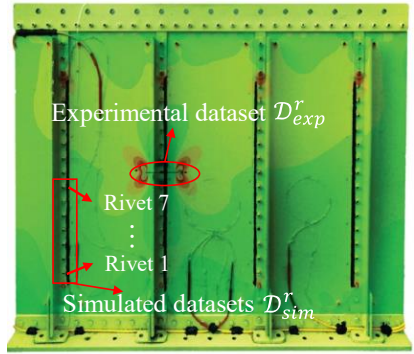
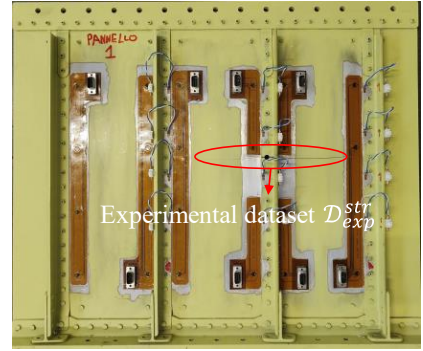

(a) (b)

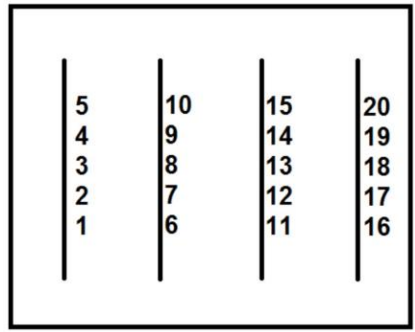
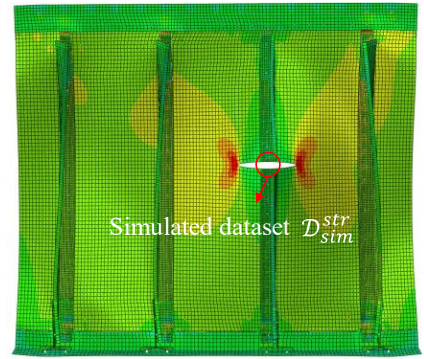

(c) (d)

Figure 4 Aluminum alloy stiffened panel. (a) simulated and experimental dataset for the rivet cracking damage. (b) experimental dataset for the stringer failure damage. (c) placement of 20 strain sensors. (d) simulated dataset for the stringer failure damage.

### 4.2. Dataset of the rivet cracking and stringer failure

For the rivet cracking, the crack at different rivet locations would lead to various strain distributions. Thus, the responses of strain sensors to damage size (crack length) would also exhibit differences, which is confirmed by the comparison of simulated strains shown in Figure 6. In Figure 6, the absolute strain at sensors 1, 5, and 6 and the maximum value among 20 sensors are shown, from which we can see that the relationship between the strains and the crack length varies when cracks are located at different rivets. Therefore, domain adaptation is desired if the data from one rivet is utilized to assist the monitoring of another rivet.

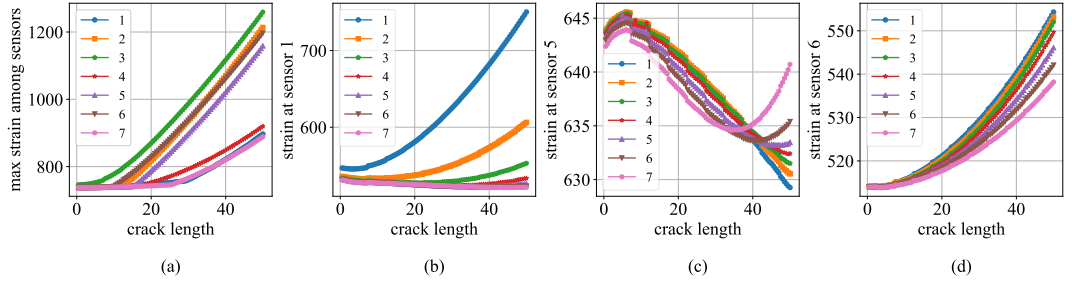

(a) (b) (c) (d)

Figure 5 The relationship of strains and crack lengths for different rivets (by simulation). From left to right are (a) the maximum strains among sensors, (b) strains at sensor 1, (c) strains at sensor 5, and (d) strains at sensor 6 respectively. 1,2, … ,7 represent the number of the rivet shown in Figure 4(a).

Here two types of datasets are considered.

In the first, the numerical model is used to generate simulated datasets at different rivets, consisting of 100 samples each. The labels are deterministically sampled ranging from $0.5\ mm$ to $50\ mm$ with step $0.5\ mm$ for rivet 1 to rivet 7, meaning the distribution of labels in different domains is the same. The seven datasets with equal spacing of crack length are represented as $\mathcal{D}_{es}^1, \mathcal{D}_{es}^2, \ldots, \mathcal{D}_{es}^7$.

In the second, the crack growth of a rivet at the second stringer, as shown in Figure 5(a), is monitored. The strains from 20 sensors and corresponding crack lengths are recorded in an experimental dataset, consisting of 800 samples. The minimum crack length is 16.53 mm, as any crack length smaller than this value is not recorded in the experiment. It is worth mentioning that the spacing of these samples is not equal and many of them are concentrated in the region where crack length is less than 20mm. The distribution of the crack lengths is shown in Figure 6. A corresponding simulated dataset is also generated. A sensor network calibration on the FE model [37] is conducted to reduce the biases due to manufacturing tolerances of the specimen, sensor positioning and installation. Furthermore, since the load significantly affects the absolute value of strain during the experiment, as in [37], a damage index is adopted here as the extracted feature to remove the effect of load. Assuming that the $N$ strain sensors measure precisely in the same direction, the damage index for the $k$-th sensor is defined as follows.

$$\varepsilon_k^{norm} = \frac{\varepsilon_k}{\sum_{i=1}^{N} \varepsilon_i / N} \tag{12}$$

The experimental dataset and corresponding simulation dataset of the rivet cracking are represented as $\mathcal{D}_{\exp}^{r}$ and $\mathcal{D}_{\sim}^{r}$, with which the domain adaptation from simulation to experiment is conducted.

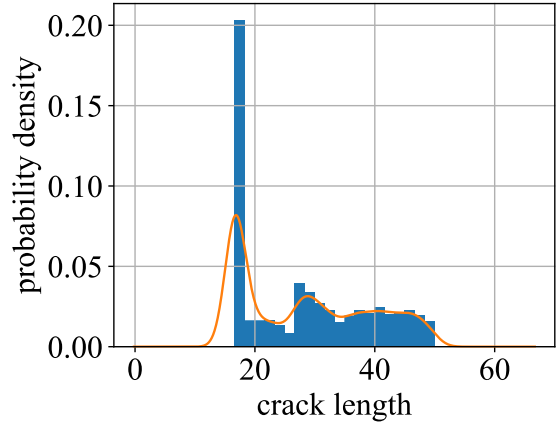

Figure 6 The distribution of crack lengths for the experimental dataset. Many samples are concentrated in the region where crack lengths less than 20. The orange lines are fitted distributions by a kernel density estimation.

Furthermore, the damage of stringer failure is considered. As with the rivet cracking dataset, the simulated and experimental datasets were both generated using the same processing method. In the experiment of the stringer failure, 13 samples are measured, from 0 mm to 197 mm. The strain response is also converted to the damage index following Eq. (12), which forms the experimental dataset, denoted as $\mathcal{D}_{exp}^{str}$. Since the maximum crack is larger than the rivet cracking, 200 simulated samples, ranging from $1\ mm$ to $200\ mm$ with step $1\ mm$, are generated, denoted as $\mathcal{D}_{sim}^{str}$.

### 4.3. Implementation and parameters setting

There are three types of domain adaptation for the damage quantification are conducted in this study.

(1) Domain adaptation across different damage locations with the same damage type, in which the dataset in one rivet location (as the source domain) can be chosen to assist the online damage quantification in another rivet location (as the target domain). Seven simulated datasets of rivet cracking with equispaced crack lengths, i.e., $\mathcal{D}_{es}^{1}, \mathcal{D}_{es}^{2}, \ldots, \mathcal{D}_{es}^{7}$, are adopted as seven domains.

(2) Domain adaptation from simulation to experiment, in which the simulation dataset is the source domain, and the experimental domain is the target domain. The rivet cracking and stringer failure are both considered. That is, domain adaptation from $\mathcal{D}_{sim}^{r}$ to $\mathcal{D}_{exp}^{r}$, and from $\mathcal{D}_{sim}^{str}$ to $\mathcal{D}_{exp}^{str}$, respectively.

(3) Domain adaptation from one type of damage to another. In this study, the rivet cracking experimental dataset $\mathcal{D}_{exp}^{r}$ is considered as the source domain, while the stringer failure experimental dataset $\mathcal{D}_{exp}^{str}$ is considered as the target domain.

The Gaussian process regression (GPR) [39], which is widely used [40,41] and works well on small datasets, is chosen as the regressor for all tasks.

For the online damage quantification task, one of these datasets is chosen as the source domain $\mathcal{D}^s$, while another dataset is then chosen as the target domain $\mathcal{D}^t$. At the beginning of the task, there are $n_{tl}^0$ labeled samples, denoted as $\mathcal{D}^{tl}$. At each step, $n_{tu}$ unlabeled samples, denoted as $\mathcal{D}^{tu}$, are processed, the domain adaption is conducted, and their labels are predicted. Then the labels of the $n_{tu}$ samples are revealed and added to the $\mathcal{D}^{tl}$. Here the $n_{tu}$ is also referred to as the prediction step-size $\Delta N$. In other words, $n_{tl} = n_{tl}^0, n_{tl}^0 + \Delta N, n_{tl}^0 + 2\Delta N, \dots, n_t$ respectively during the online damage quantification. It is worth noting that the crack lengths in $\mathcal{D}^{tl}$ are smaller than those in $\mathcal{D}^{tu}$ because of the nature of damage growth. $n_{tl}^0 = 5$ when the target domain is the rivet cracking dataset and $n_{tl}^0 = 1$ when it is the stringer failure dataset.

The prediction step-size $\Delta N$ determines the number of unlabeled data in each prediction, significantly affecting the adaptation quality and the prediction accuracy. For the domain adaptation across simulated datasets of the rivet cracking, three prediction step-sizes are considered to investigate the effects on the domain adaptation.

1. $\Delta N = 1$ is defined as a small step-size, which means once one unlabeled data is processed, the adaptation is conducted, and the label is predicted, after which the measurement is conducted.

2. $\Delta N = 5$ is defined as a medium step-size, which means the measurement is taken once five unlabeled data are acquired.

3. $\Delta N = 10$ is defined as a large step-size, which means that the measurement is conducted once ten unlabeled data are acquired, which is 1/10 of the lifecycle.

For the domain adaptation from simulation to experiment of the rivet cracking, $\Delta N$ are set to $5, 10, 30, 50$ due to large sample number.

For the domain adaptation from simulation to experiment of the stringer failure, and from the rivet cracking experiment to the stringer failure experiment, $\Delta N$ are set to 1 due to very few experiment samples of the stringer failure.

In operational environments, strain measurements are often affected by noise, which would disturb the distributions of features and decrease the quality of domain adaptation. Here, additive noise is added to the simulated datasets $\mathcal{D}_{es}^1, \mathcal{D}_{es}^2, \dots, \mathcal{D}_{es}^7$ as follows:

$$\bar{\boldsymbol{\varepsilon}} = \boldsymbol{\varepsilon} + \mathcal{N}(\mathbf{0}, \sigma_\varepsilon) \tag{13}$$

where $\bar{\boldsymbol{\varepsilon}}$ is a $20 \times 1$ vector representing the strains with noise. $\boldsymbol{\varepsilon}$ is also a $20 \times 1$ vector that represents the simulated strains. $\mathcal{N}$ represents the normal distribution and the standard deviation $\sigma_\varepsilon$ is defined as the noise level.

For all analyzed cases, the values of parameters of OFJDAR used for this application are shown in Table 2. The selection of $C$ and percentiles is consistent with [31]. $\lambda = 1$ means assigning the same weight to the MMD difference and the variance of the data. $k = 100$ to ensure that most features are preserved after the kernel operation.

Table 2 Parameters of the OFJDAR algorithm for the panel application.

| Parameter | Value |
|---|---|
| $C$ | 3 |
| $m$ | 20 |
| $\lambda$ | 1 |
| $k$ | 100 |
| Percentiles | $p_5, p_{50}, p_{95}$ |

### 4.4. Baseline methods

Four baseline methods are presented in this subsection to compare the performance of the proposed method.

*Only source domain (OSD)*: Only the labeled strain data in the source domain $\mathcal{D}^s$ are chosen to train the regressor, which would certainly have a bias, due to different effects of cracks at different rivets, as shown in Figure 5.

*Only target domain (OTD)*: Only the available labeled strain data $\mathcal{D}^{tl}$ (with smaller damage sizes) in the target domain (thus for the target rivet) are chosen to train the regressor, the prediction by which will consist of an extrapolation since the damage sizes of the unlabeled samples are larger than available labeled samples.

*Combine two domains (CTD)*: the labeled strain data in the source domain $\mathcal{D}^s$ and target domain $\mathcal{D}^{tl}$ of the two previous cases are directly combined to train the regressor.

*Online Transfer Component Analysis for Regression (OTCAR)*: the labeled strain data both in the source and target domain are adapted to a shared space by Algorithm 3, in which the adapted data are utilized to train the regressor. In the adaptation, OTCAR only considers the discrepancy of marginal distribution and thus does not take the samples' label into consideration.

### 4.5. Performance evaluation

The prediction accuracy is evaluated by Root Mean Square Error (RMSE) function.

$$\text{RMSE} = \sqrt{\frac{1}{n}\sum_{i=1}^{n}(\mathbf{y}_i - \hat{\mathbf{y}}_i)^2} \qquad (14)$$

During an online damage quantification task, the predicted crack lengths, using the proposed OFJDAR and four baseline methods, would be stored at each step. When all the predictions are made, the stored predictions, consisting of 95 samples for the simulated datasets and 795 samples for the experimental dataset in this study, would be compared with the true lengths by calculating the RMSE with Eq. (14).

## 5. Results and Discussions

In this section, several comparisons are carried out under various factors, including prediction step-size $\Delta N$, different domain variability, additive noise of strains, and the effect of random crack length spacing. In Section 5.1, the results of online damage quantification with domain adaptations from $\mathcal{D}_{es}^1$ to $\mathcal{D}_{es}^2$ under three prediction step-sizes are compared. Section 5.2 compares the online damage quantification with domain adaptation over different rivet locations, as different domains. In section 5.3, three noise levels with normal distribution are applied to the simulated strains, and the results are compared to simulate the algorithm performance in a more realistic environment. In Section 5.4, the domain adaptation from the simulated dataset $\mathcal{D}_{sim}$ to the experimental dataset $\mathcal{D}_{exp}$ are conducted, which is a realistic scenario in engineering. Finally, in Section 5.5, a challenging domain adaptation, from the rivet cracking experiment to the stringer failure experiment, is performed, showing the great potential of the proposed method for damage quantification in realistic environments.

### 5.1. Comparison of different prediction step-sizes.

Consider $\mathcal{D}_{es}^1$ as source domain and $\mathcal{D}_{es}^2$ as target domain. The three step-sizes defined in Section 4.3 are tested, and results are shown in Figure 7.

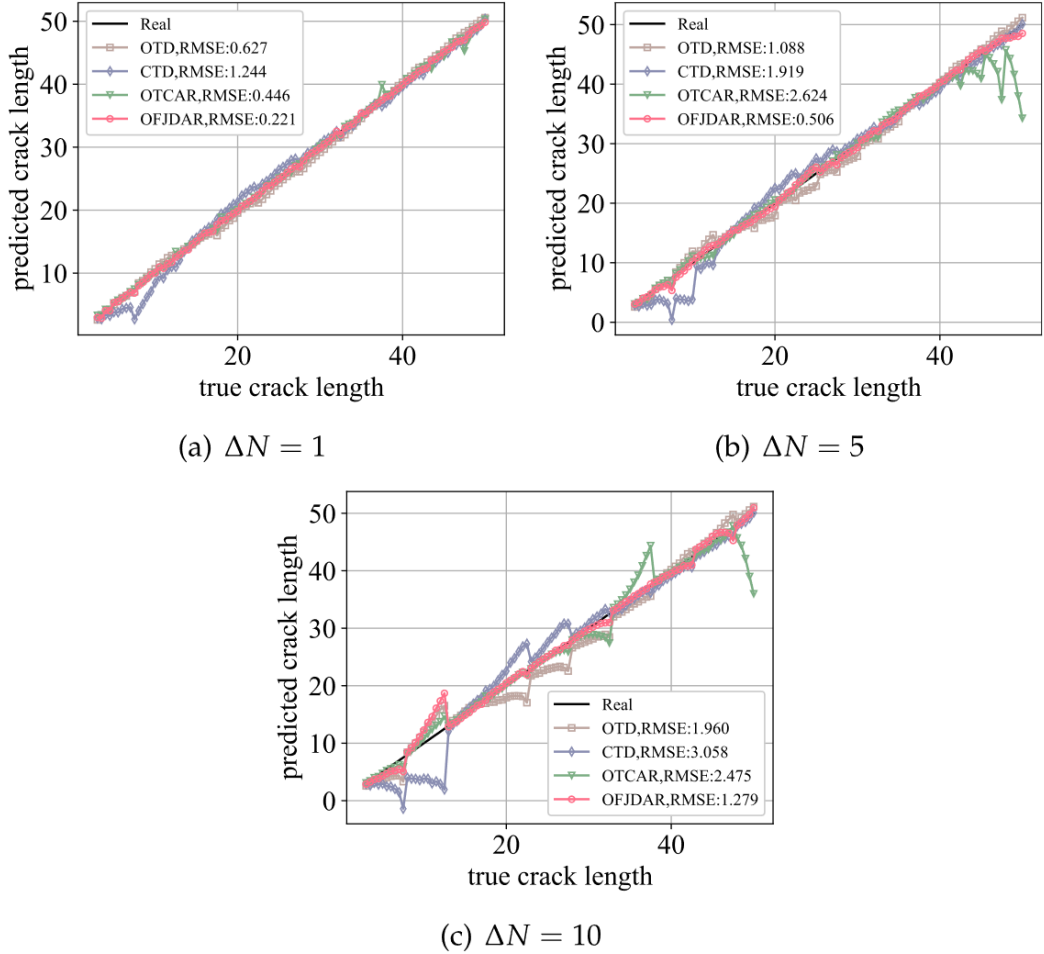

Figure 7 Effects of the prediction step-size on the prediction accuracy. Source domain: $\mathcal{D}_{es}^1$, and target domain: $\mathcal{D}_{es}^2$. (a) small step-size $\Delta N = 1$, (b) medium step-size $\Delta N = 5$, (c) large step-size $\Delta N = 10$

The RMSE of baseline methods OTD and CTD increases significantly when the prediction step-size grows. Especially for OTD, it is easily explained as with the increase of $\Delta N$, the correlation between the new samples and labeled samples gradually decreases for all the extrapolation problems. Thus, the GPR model trained on labeled samples cannot estimate unlabeled samples accurately. It is worth noting that an essential characteristic of GPR for OTD is that, after each $\Delta N$, the error grows gradually from 0. Once the true damage size is acquired and a new regressor is trained, the error is reset to 0. For CTD, because of the direct combination of the source and target domain, the distribution quality might be disturbed and sometimes lead to inferior prediction. The baseline domain adaptation method, OTCAR, has better performance than OTD and CTD at most damage sizes. However, for the medium and large step-sizes, as shown from Figures 7(b) and 7(c), the prediction of larger cracks becomes unstable, which might be due to OTCAR not learning a suitable transformation when the difference of conditional distribution grows. However, OTCAR, which only considers the marginal contribution discrepancy, could improve the accuracy of the damage quantification after the adaptation.

The RMSE of OFJDAR is smaller than all baseline methods. When the prediction step-size is larger, its superiority compared with OTD and CTD is decreased but can still hold the best performance. Results show that, by also considering the conditional distribution discrepancy, the proposed OFJDAR can achieve better domain adaptation than OTCAR, thus improving the accuracy of online damage quantification.

## 5.2. Comparison of prediction accuracy across domains

As shown in Figure 5, the crack length influence on the strain pattern varies for cracks located at different rivets, which causes a large discrepancy between domains and have a significant effect on the quality of the domain adaptation. Here, two comparisons are conducted to evaluate the prediction accuracy across domains.

First, $\mathcal{D}_{es}^1$, $\mathcal{D}_{es}^2$, and $\mathcal{D}_{es}^3$ are chosen as three domains, and the cross adaptations form six domain adaptation scenarios. The prediction accuracies of the proposed and baseline methods are listed in Table 3, exclusively for $\Delta N = 5$.

Table 3 Accuracy (RMSE) of six cross-domain adaptations on rivet-crack datasets

| Domain | OSD | OTD | CTD | OTCAR | OFJDAR |
|---|---|---|---|---|---|
| $\mathcal{D}_{es}^1 \to \mathcal{D}_{es}^2$ | 18.988 | 1.088 | 1.919 | 2.454 | **0.506** |
| $\mathcal{D}_{es}^1 \to \mathcal{D}_{es}^3$ | 27.627 | 0.962 | 1.222 | 1.834 | **0.899** |
| $\mathcal{D}_{es}^1 \to \mathcal{D}_{es}^3$ | 76.788 | 0.974 | 0.956 | 1.337 | **0.598** |
| $\mathcal{D}_{es}^2 \to \mathcal{D}_{es}^1$ | 106.914 | 0.882 | 1.509 | 0.852 | **0.621** |
| $\mathcal{D}_{es}^3 \to \mathcal{D}_{es}^2$ | 19.802 | 0.910 | 0.865 | 2.914 | **0.465** |
| $\mathcal{D}_{es}^3 \to \mathcal{D}_{es}^1$ | 66.720 | 0.851 | 1.188 | 1.439 | **0.822** |

As shown in Table 3, OSD has a large RMSE, which proves that directly applying the model trained in the source domain to data in the target domain is unfeasible. For methods that do not consider the domain adaptation, CTD predicts worse than OTD in most cases, confirming that directly combining two domains might not be a feasible solution. Our proposed method, OFJDAR, has the best performance in all the six adaptations scenarios. Interestingly, in the adaptation from $\mathcal{D}_{es}^1 \to \mathcal{D}_{es}^3$, the RMSE of OTD is slightly higher than that of OFJDAR, while in the reverse adaptation $\mathcal{D}_{es}^3 \to \mathcal{D}_{es}^1$, the RMSE of OTD is decreased to a lower level than OFJDAR. From Figure 3 we can see that Rivet 1 is closer to sensor 1 than rivet 3, thus a crack in rivet 1 would have a higher sensitivity at the position of sensor 1. Consequently, the adaptation quality in $\mathcal{D}_{es}^1 \to \mathcal{D}_{es}^3$ results higher than that in $\mathcal{D}_{es}^3 \to \mathcal{D}_{es}^1$.

In the second comparison, $\mathcal{D}_{es}^1$ (crack at rivet 1) is set as the source domain, while $\mathcal{D}_{es}^2$ to $\mathcal{D}_{es}^7$ (cracks in rivets 2 to 7) are chosen as the six target domains. The prediction accuracy under different steps is compared in Figure 8.

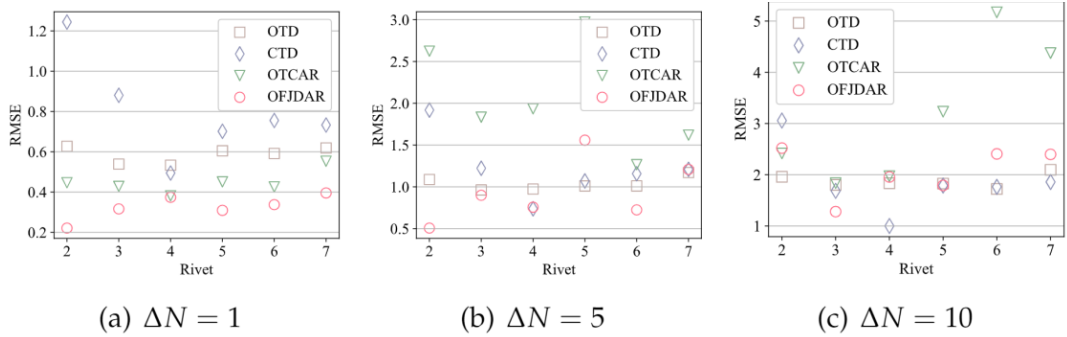

(a) $\Delta N = 1$  (b) $\Delta N = 5$  (c) $\Delta N = 10$

Figure 8 Comparison of the prediction accuracy across different domains. $\mathcal{D}_{es}^1$ (at rivet 1) is the source domain, and $\mathcal{D}_{es}^2$ to $\mathcal{D}_{es}^7$ (from rivet 2 to 7) are target domains.

Overall, the performance of OTD remains stable across domains, whereas that of CTD displays significant variability. A reasonable explanation is that the direct combination raises significant uncertainties, lowering the prediction accuracy. In most circumstances, the proposed methods show the best performance and outperform OTCAR, and the latter shows large fluctuations as the prediction step-size grows.

The proposed methods show the best accuracy in all target domains when the step is small, where the adaptation quality could be guaranteed as the effects of pseudo labels are slight. The accuracy of OTCAR is less than the proposed method but still outperforms other baseline methods without adaptation. When the step is medium, the proposed method still has the best performance except at $\mathcal{D}_{es}^5$ (rivet 5), while the accuracy of the OTCAR shows notable fluctuation.

## 5.3. Additive noise in strain measurement

Results with three tested noise levels are shown in Figure 9, specifically when $\Delta N = 5$. With the noise level rising, the adaptation quality is markedly decreased because the additive noise undermines the relevance of samples and labels. However, the proposed method outperforms baseline methods up to $\sigma_\varepsilon = 10$ με [1], demonstrating to be robust to a noisy environment. Noticeably, the accuracy of OTCAR drops with the increase in the noise level, demonstrating that only considering the marginal distribution would be insufficient in a noisy environment.

---

[1] Selected as a typical measurement noise level for realistic applications involving strain gauges.

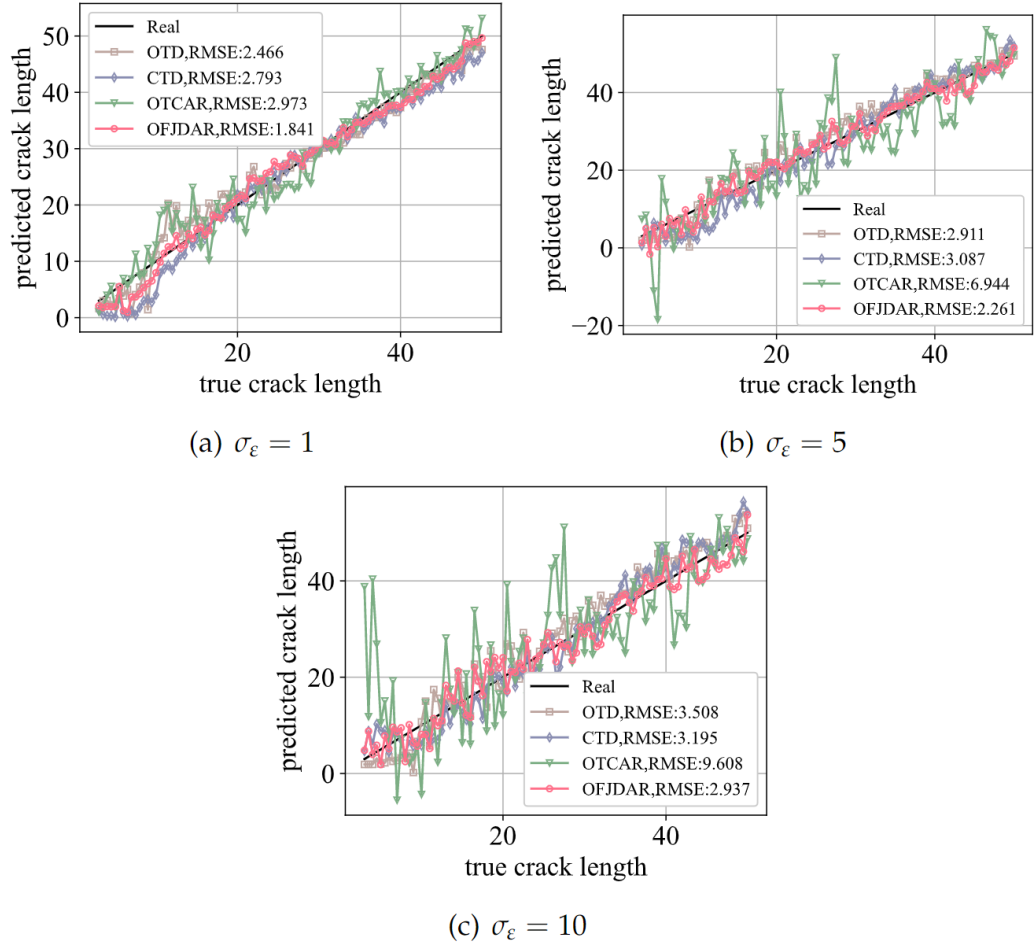

Figure 9 Effects of different noise levels on the prediction accuracy. $\Delta N = 5$

## 5.4. Adaptation from simulation to experiment

It is a common scenario for the digital twin that the damage quantification model is off-line trained by simulations and deployed online to quantify the damage state with the real SHM data. For the simulated dataset, there are inevitably biases compared to the realistic structure. Meanwhile, for the experimental dataset, the influences of environmental and operational variability (EOV) conditions are also significant [35,42,43]. Due to a variety of factors, including measurement noise, unidentified ambient excitations, environmental variability, and others, bias and variance are introduced into the dataset. In this case, the simulated dataset $\mathcal{D}_{sim}$ is selected as the source domain, while the experimental dataset $\mathcal{D}_{exp}$ is the target domain.

Figure 10 shows the domain adaptation for the rivet cracking damage, the predictions by the OTD and CTD have large biases in the 20~30 mm, and the biases increase with the step-size. This is because the measurement of strain is noisy in the experiment, making the relationship between the damage index and the crack length not monotonic. In the second half of the process, the correlation between strain and damage becomes more prominent as the crack length increases, thus the quality of the prediction is better. It should be noted that, the error of the OSD method is significant, as shown in Table 4, indicating that quantifying the damage during the experiments directly using the model trained on the simulation data may not be feasible if the simulation is biased and the experiment is noisy.

The proposed OFJDAR shows the best performance in the presented four step-sizes, especially in the 20~30 mm, where OTD and CTD have poor predictions. However, at the beginning of crack propagation and for several small prediction steps, OFJDAR shows significant fluctuation, which is even more severe in OTCAR. It can be explained as the presence of noise misdirects the adaptation. Nevertheless, in most of the processes, the domain adaptation by the proposed method improves the accuracy of quantification.

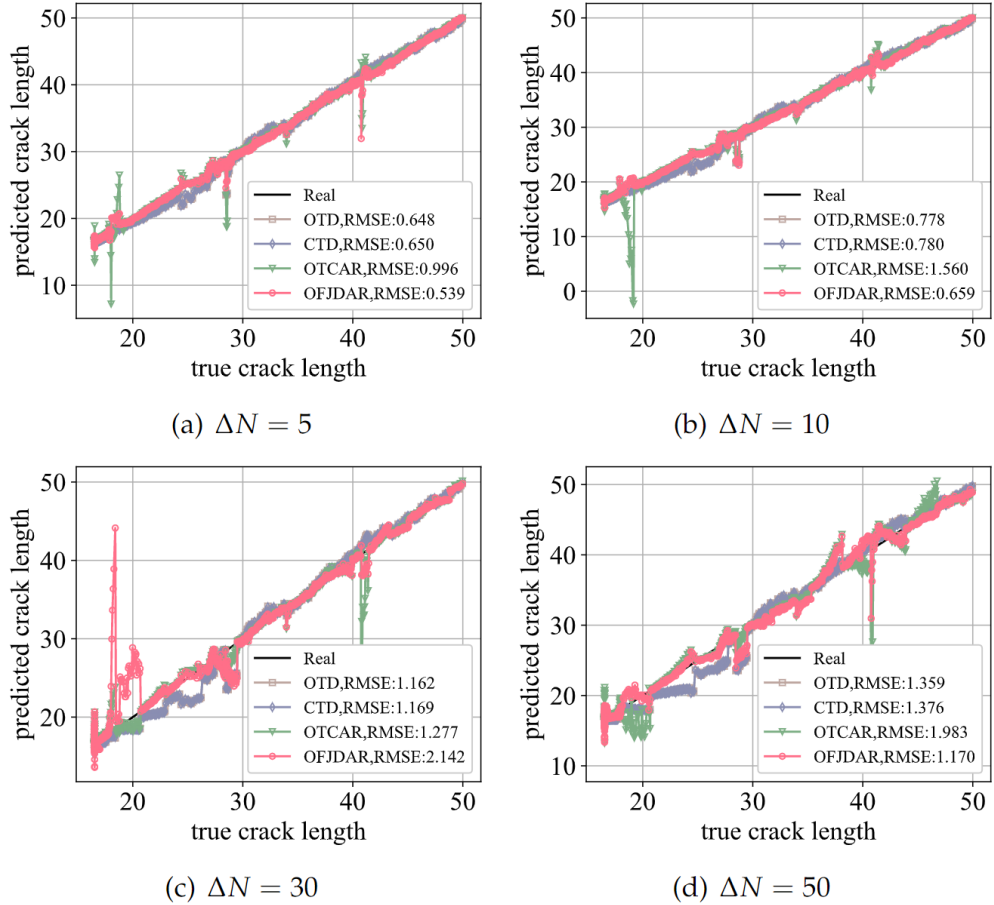

Figure 10 Effects of different step-sizes on the prediction accuracy of domain adaptation from simulation to experiment with the rivet cracking damage.

Table 4 Accuracy (RMSE) of domain adaptations from simulation to experiment.

| $\Delta N$ | OSD | OTD | CTD | OTCAR | OFJDAR |
|---|---|---|---|---|---|
| 5 | 171.703 | 0.648 | 0.650 | 0.996 | **0.539** |
| 10 | 171.647 | 0.778 | 0.780 | 1.560 | **0.659** |
| 30 | 171.372 | 1.162 | 1.169 | 1.356 | **1.048** |
| 50 | 171.143 | 1.359 | 1.376 | 1.983 | **1.170** |

The result of the adaptation for the rivet cracking damage is shown in Figure 11. Due to the larger crack size compared to that of the rivet cracking, the overall RMSE also increases. However, the method proposed in this paper still has the best performance. In addition, it can be observed that at the last sample point, the error of the method without applying domain adaptation is very large.

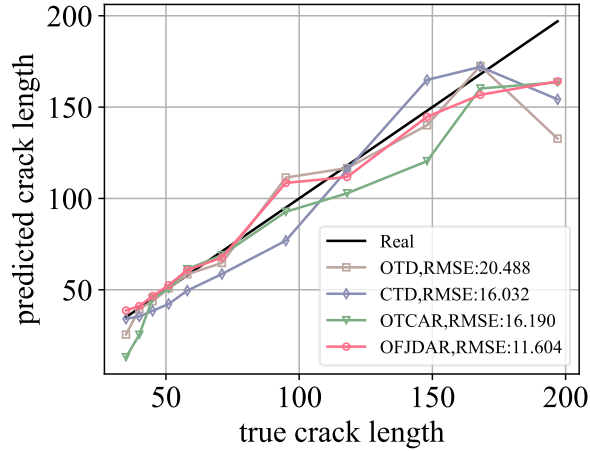

Figure 11 Domain adaptation from simulation to experiment with the stringer failure damage

## 5.5. Adaptation from the rivet cracking experiment to the stringer failure experiment

Compared to rivet cracking, string failure is a more serious damage to the structural integrity. Here the use of existing experimental data on rivet damage to enhance the accuracy of damage quantification for string bar failure is explored. Since the maximum crack length of the rivet cracking dataset was only 95 mm, the number of samples in the stringer failure dataset is reduced to 9, making the maximum length 97 mm. It is worth noting that the two damages are located at different locations, on the second and third stringers. Although the strain sensor responses are symmetrical to a certain degree due to the symmetry of the structure, it still adds more difficulty to the domain adaptation.

Figure 12 reports the results of domain adaptations with different methods. It can be seen that the proposed OFJDAR also outperforms the baselines, with an improvement of nearly 50%. There is not much difference between the CTD and OTCAR, which both have large errors. It may be because the marginal distribution of the source and target domains is similar, making the TCA not very effective.

During the service stage, structures may present various types of damage. In the face of new damage types, using the method proposed in this paper, we can leverage the data of prior similar damages to improve the damage accuracy of current damage type, which will greatly expand the availability of digital twins.

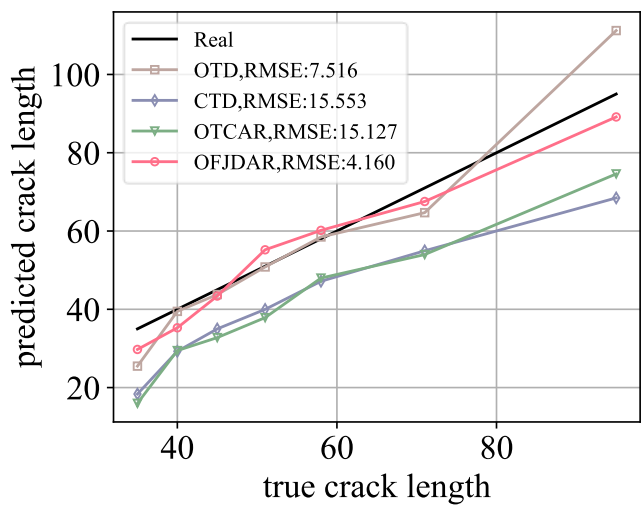

Figure 12 Domain adaptation from the rivet cracking experiment to the stringer failure experiment. $\Delta N = 1$.

## 6. Concluding remarks

A novel domain adaptation method for regression, OFJDAR, is proposed in this study. The main contribution of this method is to simultaneously consider the marginal and conditional distribution adaptions for regression problems by converting the continuous real-valued labels to fuzzy class labels in the computation of MMK matrices, thus forming a modified JDA method for regression, which was not well addressed in previous studies.

Based on the proposed method, a framework for online structural damage quantification with domain adaptation is suggested. The proposed domain adaptation method is coupled with an automatic hypermeter optimization strategy. Thus, knowledge from similar structures/damages could be adapted to assist the damage quantification in the current structure.

A helicopter panel case with two types of damage (rivet cracking and stringer failure), is used to demonstrate the proposed framework. Three types of domain adaptations are conducted. The first one considers the adaptation across various locations, which is common when damage initiates from different critical locations in the same structure. The second one used the simulated dataset from the digital twin to assist the damage quantification on the realistic structures, and the third one carries out the adaptation from a prior damage type to a new damage type. The proposed method is benchmarked with various state-of-the-art methods that account for several factors, including prediction step-size, different domain variability, and additive noise. Results demonstrate that the proposed domain adaptation methods and framework could significantly improve the online damage quantification accuracy in all three types of adaptations, and also perform robustly in a realistic environment under environmental and operational variability (EOV) conditions.

The digital twin is the fact of coupling the model with sensor data to mimic the real structure on a computer and then enabling the prognosis, thus supporting the decision-making process. Considering the increase of digital twin development from individual tracking [6,44] to fleet management[45], one cannot expect a model fitted on a fleet component to perform perfectly well with no discrepancy on a second component. Some researchers studied the feasibility of applying domain adaptation in digital twins [7,46], but practical methods are still insufficient. In our future works, by combining the proposed framework with a physics-based reduced-order model [41], domain adaptation can improve the accuracy of model updating and structural prognosis in this framework.

## CRediT authorship contribution statement

**Xuan Zhou**: Conceptualization, Methodology, Software, Formal analysis, Investigation, Writing - original draft. **Claudio Sbarufatti**: Methodology, Curation, Writing - review & editing,. **Marco Giglio**: Writing - review & editing, Funding acquisition. **Leiting Dong**: Methodology, Writing - review & editing, Funding acquisition.

## Declaration of Competing Interest

The authors declare that they have no known competing financial interests or personal relationships that could have appeared to influence the work reported in this paper.

## Acknowledgment

The authors would like to acknowledge the support of the Aeronautical Science Foundation of China (grant number 201909051001) and the China Scholarship Council (grant number 202106020002).